\documentclass[%
 reprint,
superscriptaddress,
 amsmath,amssymb,
 pra,
floatfix,
]{revtex4-2}

\usepackage{hyperref}
\usepackage{graphicx}%
\usepackage{multirow}%
\usepackage{amsfonts}%
\usepackage{amsthm}%
\usepackage{mathrsfs}%
\usepackage[title]{appendix}%
\usepackage{xcolor}%
\usepackage{textcomp}%
\usepackage{booktabs}%
\usepackage{algorithm}%
\usepackage{algorithmicx}%
\usepackage{algpseudocode}%
\usepackage{listings}%
\usepackage{braket}%
\usepackage{bm}
\usepackage[normalem]{ulem}%

\newtheorem{theorem}{Theorem}%  meant for continuous numbers
\newtheorem{proposition}[theorem]{Proposition}% 

\begin{document}

\title{Nonlinearity and Quantum Metrology in the Double-Morse Potential}

\author{Firoz Chogle}
\affiliation{Department of Applied Mathematics and Sciences, Khalifa University, Abu Dhabi, UAE}

\author{Berihu Teklu}
\affiliation{Department of Applied Mathematics and Sciences, Khalifa University, Abu Dhabi, UAE}
\affiliation{College of Computing and Mathematical Sciences, Center for Cyber-Physical Systems (C2PS), Khalifa University of Science and Technology, Abu Dhabi, UAE}

\author{Jorge Zubelli}
\affiliation{Department of Physics, Khalifa University, Abu Dhabi, UAE}

\author{Ernesto Damiani}
\affiliation{College of Computing and Mathematical Sciences, Center for Cyber-Physical Systems (C2PS), Khalifa University of Science and Technology, Abu Dhabi, UAE}
\affiliation{Dipartimento di Informatica, Universit{\`a} degli Studi di Milano, Via Giovanni Celoria 18, Milano, 20133, Italy}


\begin{abstract}
    We address the nonlinear properties of the double-Morse potential as a resource for single-mode quantum states due to its double-well structure and anharmonicity. 
    We obtain analytical expressions for the ground-state wavefunction and the corresponding ground-state energy, using the asymmetry (width) parameter $\alpha$ as the primary control parameter. 
    We then assess non-Gaussianity and nonclassicality as quantitative signatures of nonlinearity and quantumness, and we find that both increase monotonically with $\alpha$. 
    Furthermore, we analyze the metrological performance of the model for estimating the structural parameter $\alpha$. By evaluating the corresponding Fisher information, we show that position 
    measurements are optimal and can saturate the Cram\'er--Rao bound. In particular, the estimation of $\alpha$ is most precise in the shallow-well regime, where the quantum Fisher information is largest. 
    For deep wells, enhanced sensitivity is instead obtained for the reparameterized control variable $A=2e^{-\alpha x_0}$, provided that $x_0$ is independently calibrated. 
    These results establish the double-Morse potential as a controllable source of non-Gaussianity and nonclassicality, with a metrological behavior that depends on the chosen estimation parameter. 
    We highlight possible applications of this model in quantum sensing, continuous-variable quantum information, and quantum simulation.
\end{abstract}

\maketitle

%\tableofcontents

\section{Introduction}\label{sec1}

Quantum technologies exploit intrinsically quantum features, including superposition, entanglement, and phase-space nonclassicality, to enable applications in computation \cite{shahandeh2019quantum}, secure communication \cite{BB84, Ekert1991,Bennett1992dense, Caves1994}, sensing and metrology \cite{Yuen1993, Matteoprl, Teklu_2009, PhysRevLett.121.243601, Candeloro, asjad2023joint, XIE2023106575, Huang2024,Kwon2019PRL, Chin2012}, and machine learning \cite{PhysRevApplied.15.044003, Ilaria22}. The concept of nonclassicality first attracted significant attention in quantum optics, where phenomena such as photon antibunching, squeezing, and negativity of quasiprobability distributions provided operational signatures of behavior beyond classical wave descriptions. In this context, single-mode bosonic systems provide a natural and experimentally accessible platform for generating, manipulating, and characterizing non-Gaussian states. Identifying which quantum features are genuinely nonclassical and developing rigorous methods to quantify them has thus become a central theme in the theory and application of quantum technologies \cite{Lvovsky2015}. In particular, recent work has linked continuous-variable nonclassicality to metrological advantage and has clarified how single-mode nonclassicality may be converted into entanglement under suitable linear-optical transformations \cite{Kwon2019PRL,Killoran2016PRL}.

In recent decades, single-mode bosonic systems, such as quantized electromagnetic field modes in optical cavities or microwave resonators, have emerged as ideal candidates for the theoretical formulation and experimental investigation of nonclassicality \cite{RevModPhys.75.281, RevModPhys.86.1391,Lewenstein2013,Rogers2014,Smirne}. These systems enable precise control over the generation and characterization of quantum features \cite{Peano_2006,Ong,DiVincenzo_2012,Simon,Vacanti,Victor}. Moreover, nonlinear interaction models accessible through these platforms are experimentally realizable in various oscillatory systems, including mechanical and electromagnetic ones \cite{Home_2011,Sankey2010}. Nevertheless, a comprehensive understanding of how nonlinearity influences quantumness remains an open and largely unexplored question.

The quantum-to-classical transition in nonlinear nanoelectromechanical systems (NEMS) has been examined in previous studies, including those conducted by Katz et al. \cite{Katz,Katz_2008}, which considered both closed and open system dynamics. These investigations examined how quantum signatures appear and are suppressed, particularly as thermal effects increase. Based on this framework, more recent research has started to investigate the structural function of nonlinearity as a tool for producing and maintaining nonclassicality. Teklu et al. \cite{Teklu} specifically studied a Duffing-type nonlinear nanomechanical resonator. They demonstrated that the Wigner function's negativity, a quantitative indicator of nonclassicality, rises with the system's nonlinearity. The idea that nonlinearity might be used to magnify quantum features in bosonic systems was further supported by complementary results in \cite{Francesco}, where a broader class of anharmonic potentials was studied. Despite these promising results, a general framework connecting nonlinearity to nonclassicality in the system remains an open research direction, particularly beyond specific models, such as the Duffing oscillator. Motivated by these results, we contribute to this line of inquiry by investigating a quantum double-Morse oscillator, aiming to clarify whether such a nonlinear potential can also serve as a platform for enhancing quantum effects in a tunable and structurally generalizable way. Finally, we employ techniques from local quantum estimating theory to characterize the double-Morse potential and determine the optimal approach to infer the width parameter \cite{Holevo2011,Helstrom1969,Paris2009,Albarelli2020,Chabane}. In this procedure, the figure of merit is the quantum Fisher information, which quantifies the amount of information about a parameter that can be extracted through measurements performed on a family of quantum states.\\

This paper is organized as follows: In Section~\ref {sec2} we introduce the double-Morse (DM) oscillator and fix our notation, derive the exact ground state and its spectrum in dimensionless units, and discuss how the width/asymmetry parameter $\alpha$ controls the potential’s nonlinearity. We then define and compute two complementary indicators of “resourcefulness”: a Bures-distance–based nonlinearity measure and the relative–entropy non-Gaussianity, and we obtain a closed-form expression for the ground–state Wigner function-written in terms of modified Bessel functions of imaginary order—from which we evaluate Wigner negativity and the associated nonclassicality. We also recall the \emph{entanglement potential} operational test by interfering the state with the vacuum at a $50{:}50$ beam splitter. Next, in Section~\ref{sec3}, we cast the estimation of $\alpha$ within local quantum estimation theory, review classical and quantum Fisher information (via the symmetric logarithmic derivative), and derive a compact formula for the QFI of the DM ground state, showing that position measurements saturate the quantum Cramér–Rao bound. Section~\ref{sec4} summarizes the main findings and outlines implications for quantum technologies.

\section{Preliminaries}
\label{sec2}
This section reviews the primary tools employed in the present work: the concept of the double-Morse oscillator, the quantification of nonclassicality of quantum states and nonlinearity of potentials. We begin with an overview of the double-Morse potential, followed by a discussion of its relevance in the context of quantum nonlinearity.

\subsection{Double-Morse Oscillator}

The Morse potential is commonly used to model the interaction between atoms in a diatomic molecule. It offers a more realistic model than the harmonic oscillator due to its anharmonicity and asymptotic behavior  \cite{morse1929diatomic}. It is typically expressed as
 \begin{equation}
 V_{\text{M}}(x) = D\left( e^{-2\alpha x} - 2e^{-\alpha x} \right),
 \label{eqn1}
 \end{equation}
 where $D$ denotes the bond dissociation energy, and the parameter $\alpha$ is responsible for the width of the potential well. The double-Morse potential is constructed by superimposing two Morse potentials centered at $\pm x_{0}$, each facing the other, i.e., with $\pm\alpha$. The resulting expression is
 
 \begin{equation}
 V_{\text{DM}}(x) = D\left(A\cosh{(\alpha x)} -1\right)^2,
 \label{eqn2}
 \end{equation}
 where $A=2e^{-\alpha x_{0}}$.  A constant term resulting from the summation is dropped, as it merely shifts the potential energy by a constant and does not affect the physical behavior of the system. For $0<A<1$, Eq.~\eqref{eqn2} describes a symmetric bistable potential with two wells. It is instructive to examine the nonlinear potential-energy landscape that governs the oscillator, treating $D$, $\alpha$, and $x_0$ as independent classical parameters at this stage.

We stress that this classical illustration is independent of the quasi-exact solvability (QES) condition derived below: here $D$ and $x_0$ are held fixed purely to visualize how the \emph{shape} of the potential landscape depends on $\alpha$, whereas the exact ground-state wavefunction requires $D$ and $\alpha$ to be co-varied according to $D(\alpha)=\hbar^2\alpha^2/(8m)$ [Eq.~\eqref{eqn7new2}]. The two roles of $\alpha$ --- as a free shape parameter of the classical potential (this paragraph) and as the single QES control parameter of the quantum ground state (from Eq.~\eqref{eqn7new} onward) --- should not be conflated; we return to this point explicitly once the ground state is derived.

\begin{figure}[h]
    \centering
    \includegraphics[width=0.8\columnwidth]{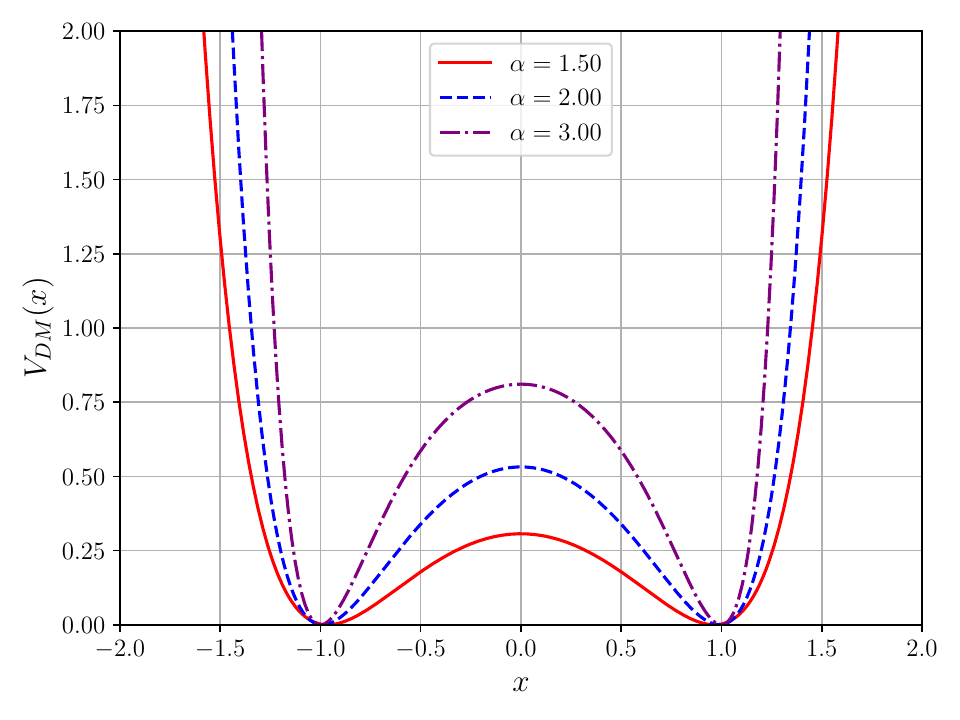}
    \caption{Double Morse potential profiles for various values of $\alpha$.}
    \label{fig1}
\end{figure}

As shown in Fig.~\ref{fig1}, when $D$ and $x_{0}$ are held fixed (so that $A=2e^{-\alpha x_{0}}$ varies with $\alpha$), increasing $\alpha$ beyond the threshold $\alpha x_{0}=\ln 2$ (i.e., $A<1$) transforms the profile from a single, nearly harmonic well into a symmetric double well. The minima move away from the origin toward $\pm x_{0}$, so the interwell separation $2|x_{\min}|$ grows from zero and approaches $2x_{0}$ as $\alpha\to\infty$, while the central barrier height $V(0)=D\big(1-2e^{-\alpha x_{0}}\big)^{2}$ rises monotonically toward $D$.

 The double-Morse potential has been extensively studied over the past four decades~\cite{razavy1980exactly,matsushita1982note,robertson1981analysis,konwent1986one,konwent1998some,zaslavskii1984new,goryainov2012model,davis1973analytical,berblinger1988double,carpenter2018dynamics,pettitt1998morse,konwent1995certain,li2013quantum}. It belongs to the class of quasi-exactly solvable (QES) models, for which analytic solutions to the Schrödinger equation exist for specific parameter choices ~\cite{ushveridze2017quasi}. 
Alternative forms of QES bistable potentials have been proposed ~\cite{razavy1980exactly,matsushita1982note,robertson1981analysis,zaslavskii1984new,konwent1986one,konwent1998some}. Still, these can be shown to be mathematically equivalent to the double Morse potential under suitable transformations—the solutions to the corresponding Schrödinger equation yield similar eigenfunctions and spectra.

The double-Morse potential has been applied in various physical contexts, such as modeling spin systems~\cite{zaslavskii1984new}, describing intra-molecular inversion~\cite{davis1973analytical}, and characterizing the motion and phase transitions of protons in hydrogen-oxygen bonds in ferroelectrics and solids \cite{matsushita1982note,robertson1981analysis,goryainov2012model}.  It has also been used to investigate classical dynamics~\cite{carpenter2018dynamics} and chaotic behavior~\cite{berblinger1988double}. More recently, quantum wave packet dynamics of the double-Morse oscillator have been studied to explore resonance phenomena and wave revivals, with potential applications in quantum information processing and computation \cite{li2013quantum,LI2015208}.

While previous work has used the double-Morse potential to model and predict experimental results, the inherent quantum features, particularly its nonclassicality and application to quantum sensing, have not been investigated in depth. Given its anharmonic nature, one can quantify the system's nonlinearity, which is closely linked to nonclassical behavior, by examining its ground state properties \cite{paris2014quantifying}.

To this end, we consider a single particle of mass $m$ under the influence of the double-Morse potential defined in Eq.~\eqref{eqn2}. The position-space wavefunction describes the probability density of finding the particle at a given position $\psi(x)$, which satisfies the time-independent Schr\"{o}dinger equation:

\begin{equation}
    -\frac{\hbar^2}{2m}\frac{\partial^2 \psi(x)}{\partial x^2} + D(A\cosh{(\alpha x)}-1)^2\psi(x) = E\psi(x) 
    \label{eqn3}
\end{equation}
where $E$ is the energy eigenvalue of the particle corresponding to the state $\psi(x)$. To simplify the analysis, we adopt dimensionless variables such that distances and energies are measured in units of $1/\alpha$ and $\hbar^2\alpha^2/(2m)$, respectively. Following the approach in Ref.~\cite{konwent1986one}, Eq.~\eqref{eqn3}  is non-dimensionalized by introducing the following transformations
\begin{equation*}
    \alpha x = 2y,\ \mu^2 = \frac{8mD}{\hbar^2\alpha^2},\text{ and } \epsilon = \frac{8mE}{\hbar^2\alpha^2}. 
\end{equation*}
The Schr\"{o}dinger equation in these new units becomes
\begin{equation}  \label{eqn4}
    \frac{d^2\psi}{dy^2} + [\epsilon - \mu^2(A\cosh{2y}-1)^2]\psi = 0.
\end{equation}
The bound-state solutions of Eq.~\eqref{eqn4} vanish asymptotically, i.e., $\psi(y) \to 0$ as $y \to \pm \infty$. This leads to the following proposition:\\
 
\begin{proposition}
    The asymptotic solution of Eq.~\eqref{eqn4} as $y \to \pm \infty$ is given by
    \begin{equation}
    \psi(y) \approx \exp\left(-\frac{\mu A}{2}\cosh{2y}\right)
    \label{eqn5}
\end{equation}
\end{proposition}
% \noindent\textbf{Proposition 1.} \textit{The asymptotic solution of Eq.~\eqref{eqn4} as $y \to \pm \infty$ is given by}
% \begin{equation}
%     \psi(y) \approx \exp\left(-\frac{\mu A}{2}\cosh{2y}\right)
%     \label{eqn5}
% \end{equation}
\begin{figure}[h]
    \centering
    \includegraphics[width=0.8\columnwidth]{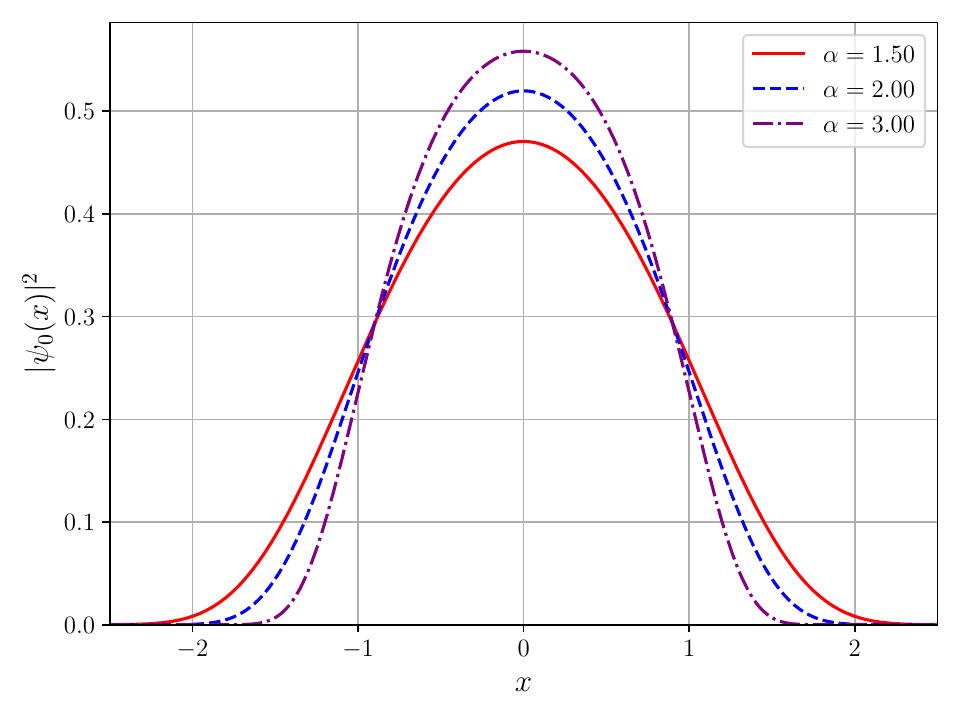}
    \caption{Ground-state probability density $|\psi_0(x)|^2$ of the double-Morse oscillator for several values of $\alpha$. Increasing $\alpha$ narrows the distribution in $x$ and enhances localization at the well minima.}
    \label{fig2}
\end{figure}
Building on Proposition~1, the general wavefunction can be expressed as $\psi(y) = \exp\left(-\frac{\mu A}{2}\cosh{2y}\right)\phi(y)$, where $\phi(y)$ satisfies the differential equation:
\begin{equation}
    \frac{d^2\phi}{dy^2} - 2A\mu \sinh 2y \frac{d\phi}{dy} + (\Tilde{\epsilon} + 2\mu A (\mu - 1)\cosh 2y)\phi = 0.
    \label{eqn6}
\end{equation}
with $\Tilde{\epsilon} = \epsilon - \mu^2(1+A^2)$. When $\mu = n + 1$, the function $\phi(y)$ becomes a polynomial in hyperbolic functions, allowing the system to support $n + 1$ bound states. The recurrence relations for the polynomial coefficients are obtained from Eq.~\eqref{eqn6}, and the condition for non-trivial polynomial solutions is that the determinant of the associated coefficient matrix vanishes. The roots of this determinant equation yield the allowed energy levels.

Since our analysis focuses on the ground state, Eq.~\eqref{eqn6} is solved for $n = 0$. The ground-state wavefunction is given by
\begin{equation}
    \psi_0(y) = C\exp\left(-\frac{A}{2}\cosh{2y}\right) 
    \label{eqn7}
\end{equation}
where $C$ is the normalization constant. The corresponding ground-state energy is $\epsilon_0 = \mu^2(1 + A^2)$. 
    
Rescaling back to the dimensional variables, the wavefunction is given as 
\begin{equation}
    \psi_0(x) = \sqrt{\frac{\alpha}{2K_0(A)}}\exp\left(-\frac{A}{2}\cosh{\alpha x}\right). 
    \label{eqn7new}
\end{equation}
Here $K_0(A)$ is the zeroth-order modified Bessel function of the second kind. For a bistable configuration, the condition $0 < A = 2e^{-\alpha x_0} < 1$ must hold, leading to the constraint
\begin{equation*}
    \frac{\ln 2}{x_0} < \alpha < \infty
\end{equation*}
and the QES condition requires that 
\begin{equation}
    D(\alpha) = \frac{\hbar^2\alpha^2}{8m}.
    \label{eqn7new2}
\end{equation}
Therefore, once the QES condition $D(\alpha)=\hbar^2\alpha^2/(8m)$ is imposed, $\alpha$ is the only free parameter left to control the shape of the potential for fixed $x_0$, and in this restricted sense it is the sole nonlinearity parameter of the family of ground states analyzed in this work.

To make this statement quantitative, rather than merely qualitative, we introduce a normalized anharmonicity parameter built from the local Taylor expansion of $V_{\mathrm{DM}}(x)$ about a well minimum $x_{\min}$, defined by $A\cosh(\alpha x_{\min})=1$. Writing $V_{\mathrm{DM}}(x_{\min}+\delta)=\tfrac12 V''(x_{\min})\delta^2+\tfrac16V'''(x_{\min})\delta^3+\tfrac1{24}V''''(x_{\min})\delta^4+\dots$, a direct calculation gives
\begin{equation}
V''(x_{\min})=2D\alpha^2(1-A^2),\qquad V''''(x_{\min})=2D\alpha^4(7-4A^2).
\label{eqn:taylorcoeffs}
\end{equation}
Comparing the quartic term to the harmonic term on the natural length scale $\ell_0=\sqrt{\hbar/(m\omega)}$, $\omega=\sqrt{V''(x_{\min})/m}$, of the local reference oscillator, we define
\begin{equation}
\eta_{\mathrm{anh}} \equiv \frac{\hbar\,V''''(x_{\min})}{12\,\sqrt{m}\,\bigl[V''(x_{\min})\bigr]^{3/2}}.
\label{eqn:etaanh}
\end{equation}
Substituting Eq.~\eqref{eqn:taylorcoeffs} together with the QES relation $D(\alpha)=\hbar^2\alpha^2/(8m)$, all dependence on $\alpha$, $\hbar$, and $m$ cancels, leaving a function of $A$ alone,
\begin{equation}
\eta_{\mathrm{anh}}(A) = \frac{7-4A^2}{6\,(1-A^2)^{3/2}}.
\label{eqn:etaanhA}
\end{equation}

\begin{figure}[h]
    \centering
    \includegraphics[width=0.7\columnwidth]{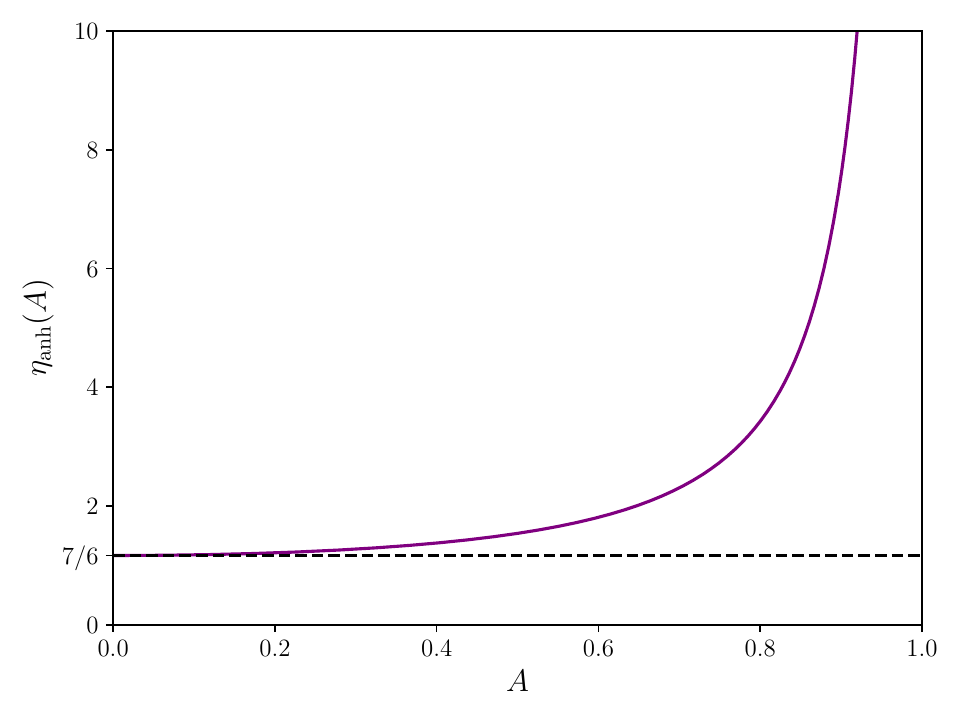}
    \caption{Normalized local anharmonicity parameter $\eta_{\mathrm{anh}}$, Eq.~\eqref{eqn:etaanhA}, as a function of $A=2e^{-\alpha x_0}$. Larger $\alpha$ (for fixed $x_0$) corresponds to smaller $A$.}
    \label{fig:etaAnh}
\end{figure}

The behavior displayed by Eq.~(12) and Fig.~3 clarifies the distinction between local anharmonicity and the global structural nonlinearity controlled by $\alpha$. 
The quantity $\eta_{\rm anh}(A)$ decreases as $\alpha$ increases, equivalently as $A$ decreases: it diverges as $A\to 1^{-}$, corresponding to the double-well 
threshold where the local curvature vanishes and the quartic contribution dominates, and approaches $\eta_{\rm anh}\to 7/6$ as $\alpha\to\infty$ ($A\to 0$). Thus, 
increasing $\alpha$ does not enhance the local, curvature-based anharmonicity of an individual well; rather, each well becomes locally closer to harmonic. The monotonic 
growth of $\eta_{\rm NG}$ and $\eta_{\rm NC}$ with $\alpha$ reported below is therefore not a local-curvature effect. Instead, it is a genuinely nonlocal structural effect, 
driven by the increasing separation and barrier height between the two wells and by the resulting quantum interference of the ground state across the barrier. In this 
sense, the relevant mechanism is a two-well delocalization effect rather than a single-well anharmonicity effect. Throughout this work, we use the term ``nonlinearity'' in this global 
structural sense, and Eq.~(12) makes explicit that this notion is distinct from local anharmonicity at the potential minimum.

Fig.~\ref{fig2} shows the ground-state probability density $|\psi_0(x)|^2$ of the double-Morse oscillator for several values of $\alpha$. As $\alpha$ increases, the density narrows in $x$ and becomes more localized around the well minima, with a correspondingly deeper dip between the wells. As we demonstrate, the ground-state wavefunction plays a central role in the analysis presented in this work. We now turn to probing the nonlinear features of the system using ground-state properties.

\subsection{Probing the nonlinear  features of a one-dimensional potential using ground states}

We begin by introducing a measure of nonlinearity that relies on the properties of the ground state (GS)~\cite{paris2014quantifying} and is constructed by comparing the perturbed ground state $\ket{\psi_0}$ with its unperturbed harmonic counterpart $\ket{0}$. Specifically, we quantify the distance between these two ground states using the Bures metric. For a perturbing nonlinear potential $\hat{V}(\hat{x})$, the nonlinearity measure $\eta_B[V]$ is defined as the appropriately normalized Bures distance $\mathcal{D}_\text{B}$ between the ground state of the oscillator under consideration and that of the corresponding harmonic oscillator. In this case, we have

\begin{equation}
    \eta_\text{B}[V] = \frac{1}{\sqrt{2}} \mathcal{D}_\text{B} \left[|\psi_0\rangle_V , |0\rangle_H   \right]
    \label{eqn8}
\end{equation}
where $|0\rangle_H$ is the GS of the reference QHO. Although this measure is intuitively appealing, identifying a suitable reference QHO potential for the given system is not feasible. The non-Gaussianity of the GS is one of the most peculiar features of a nonlinear oscillator, since it is apparent that the more the oscillator is similar to the harmonic one, the more the GS has a wave-function similar to a Gaussian. So, it is quite natural to use a bona-fide measure of non-Gaussianity for the GS to quantify the nonlinearity. In \cite{genoni2008quantifying,genoni2010quantifying}, a more robust measure of non-Gaussianity was introduced, which did not involve finding a reference potential. This approach evaluates the quantum relative entropy between the state under consideration and its reference Gaussian state. The main properties of this measure and its extension to pure states are discussed in \cite{paris2014quantifying}. For the GS of a given potential, the non-Gaussianity measure takes the form
\begin{equation}
    \eta_\text{NG}[V] = h\left( \sqrt{ \text{det}\sigma} \right)
    \label{eqn9}
\end{equation}
where $\sigma$ is the covariance matrix of the GS and the function $h(x) = \left(x+\frac{1}{2}\right) \ln\left(x+\frac{1}{2}\right) - \left(x-\frac{1}{2}\right) \ln \left(x-\frac{1}{2}\right)$. 
The covariance variance matrix calculated with the state (\ref{eqn7new}) is the following: 

\begin{equation*}
\sigma = 
\begin{pmatrix}
\langle \hat{x}^2 \rangle - \langle \hat{x} \rangle^2 & \frac{1}{2}\langle \{\hat{x},\hat{p}\} \rangle  - \langle \hat{x} \rangle \langle \hat{p} \rangle \\
\frac{1}{2}\langle \{\hat{p},\hat{x}\} \rangle  - \langle \hat{p} \rangle \langle \hat{x} \rangle & \langle \hat{p}^2 \rangle - \langle \hat{p} \rangle^2 
\end{pmatrix}.
\end{equation*}
where $\{\cdot,\cdot\}$ denotes the anti-commutator and $[\hat{x},\hat{p}]=i$. The diagonal entries are the variances of $\hat{x}$ and $\hat{p}$; symmetry implies $\sigma_{12}=\sigma_{21}$ (See Appendix \ref{covar_matrix}).

\begin{figure}[!ht]
    \centering
\includegraphics[width=0.8\columnwidth]{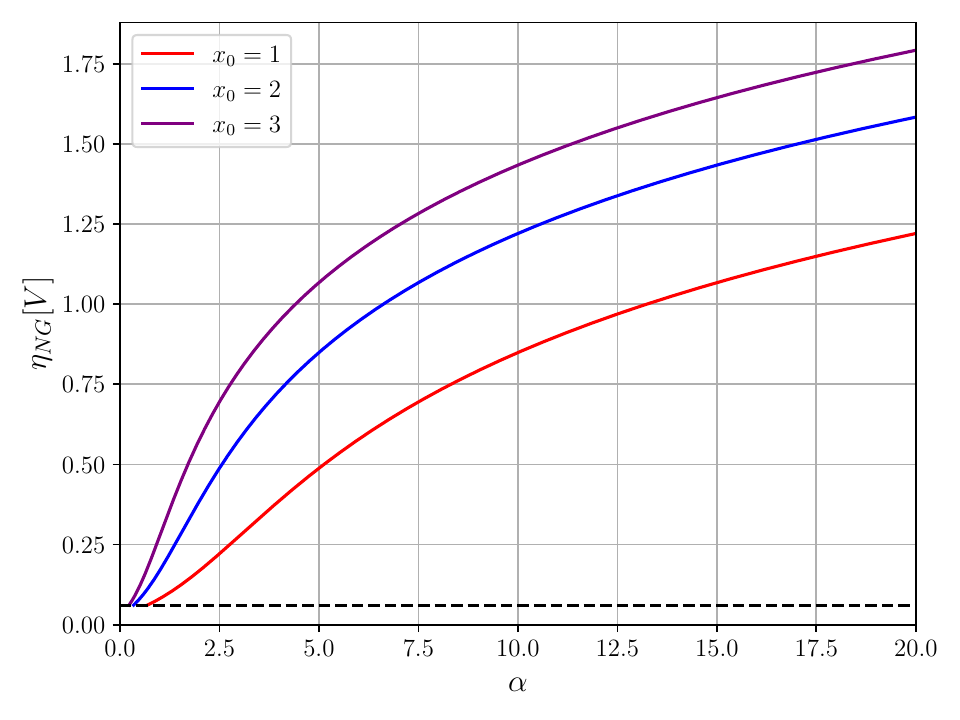}
\caption{Non-Gaussianity measure as a function of $\alpha$ for values of $x_0 = 1,2$ and $3$. The dashed black line is $\eta_{\text{NG}[V]} \approx  0.0615$. }
    \label{figNG}
\end{figure}

Fig (\ref{figNG}) shows the non-Gaussianity measure $\eta_{\mathrm{NG}}$ versus $\alpha$. 
As $\alpha$ increases, the ground state becomes more strongly non-Gaussian. 
The curves for different well separations $x_0$ illustrate that a larger $x_0$ (i.e., more widely separated minima) yields a more non-Gaussian ground state. 
To retain a double-well, we require $\alpha \in \bigl((\ln 2)/x_0,\infty\bigr)$.

The determinant of the covariance matrix, and in turn the non-Gaussianity, explicitly depend on $A(\alpha,x_0)$.
For $\alpha \approx (\ln 2)/x_0$, we have $A \approx 1$ (independent of $x_0$), so $\eta_{\mathrm{NG}}$ starts at the common value $\approx 0.0615$, indicated by the dashed line in Fig (\ref{figNG}). 
In the opposite limit, $\alpha \to \infty$ implies $A \to 0$, the covariance-matrix elements grow without bound, and the $x_0$-dependence disappears; consequently, the $\eta_{\mathrm{NG}}$ curves merge for $\alpha \gg 1$.

\subsection{Nonclassicality measure}
\label{subsec:nonclassicality}

An alternate way to quantify the quantumness of a state is to assess its deviation from a classical description via its phase-space representation, particularly the Wigner distribution. Introduced to study quantum corrections to statistical mechanics~\cite{wigner1932quantum,Cahill1967,Williams1984}, the Wigner distribution of a state $\rho$ is
\begin{equation}
    W(x,p) = \frac{1}{\pi\hslash} \int_{-\infty}^{+\infty} \!\langle x+y|\rho|x-y\rangle \,
    e^{-2ipy/\hslash}\,\mathrm{d}y .
    \label{eqn2.9}
\end{equation}
For pure states, Hudson's theorem asserts that $W(x,p)$ is everywhere non-negative if and only if the state is Gaussian~\cite{hudson1974wigner}. Equivalently, any \emph{non-Gaussian} pure state necessarily exhibits Wigner negativity. Because of these negative regions, $W$ is a \emph{quasi}probability distribution rather than a true probability density.

A common way to quantify the negativity is via
\begin{equation}
    \nu \equiv \int_{-\infty}^{+\infty}\!\int_{-\infty}^{+\infty} \bigl|W(x,p)\bigr|\,\mathrm{d}x\,\mathrm{d}p - 1,
    \label{eqn2.10}
\end{equation}
which equals twice the integrated negative ``volume,'' i.e.
$\nu = 2\!\int_{W<0}\! |W(x,p)|\,\mathrm{d}x\,\mathrm{d}p$. Following~\cite{kenfack2004negativity},  we define the (bounded) nonclassicality measure
\begin{equation}
    \eta_{\mathrm{NC}} = \frac{\nu}{\nu + 1}.
    \label{eqn2.11}
\end{equation}

Equations~\eqref{eqn2.10}--\eqref{eqn2.11} quantify \emph{Wigner-function negativity}. 
This is a widely used witness (and, in several continuous-variable settings, a resource) but it is not a universal notion of nonclassicality.
In particular, all Gaussian states have a nonnegative Wigner function, so $\eta_{\mathrm{NC}}=0$ for them; nevertheless, some Gaussian states are nonclassical in the quantum-optical sense.
A paradigmatic example is the squeezed vacuum, whose Wigner function is everywhere nonnegative despite the state being nonclassical because its Glauber--Sudarshan $P$ function fails to be a regular probability density~\cite{glauber1963coherent,sudarshan1963equivalence,walls2008quantum}.
Accordingly, in this work $\eta_{\mathrm{NC}}$ should be interpreted specifically as a measure of \emph{Wigner negativity} (and, for pure states, of departure from Gaussianity), rather than as a complete characterization of nonclassicality.
This measure vanishes for states with non-negative Wigner functions (e.g., pure Gaussian states). Note that \eqref{eqn2.11} is most meaningful for \emph{pure} states: Wigner positivity does not imply Gaussianity or classicality for mixed states. For example, incoherent mixtures of coherent states (``decohered'' Schr\"odinger-cat states) can have a positive Wigner distribution while remaining non-Gaussian~\cite{walschaers2021non}.

Using \eqref{eqn2.9}, the Wigner distribution of the ground state is
\begin{equation}
    W_0(x,p) = \frac{1}{\pi K_0(A)} \int_0^\infty e^{-A\cosh(\alpha x)\cosh y}\,\cos\left(\frac{2py}{\alpha}\right)\,\mathrm{d}y ,
    \label{eqn5.5}
\end{equation}
which can be evaluated in closed form by writing $\cos(2py/\alpha)=\cosh(2i p y/\alpha)$ and using the integral representation of the modified Bessel function \cite{abramowitz1948handbook},
\begin{equation}
    K_\nu(z)=\int_0^\infty e^{-z\cosh t}\cosh(\nu t)\,\mathrm{d}t.
    \label{eqn:handbook}
\end{equation}
This yields
\begin{equation}
    W_0(x,p) = \frac{1}{\pi K_0(A)}\, K_{(2pi/\alpha)}\!\bigl(A\cosh(\alpha x)\bigr).
    \label{eqn5.6}
\end{equation}

Here $K_{(2pi/\alpha)}$ is the modified Bessel function of the second kind (also called “third kind”) with purely imaginary order~\cite{balogh1967asymptotic,dunster1990bessel,gil2003computation,booker2013bounds}. Although numerics are typically used to evaluate $K_{(2pi/\alpha)}$, several properties are immediate:
(i) for $p\in\mathbb{R}$ and $A>0$ (double-well regime), since $\cosh(\alpha x)>0$ for all $x$, the argument $A\cosh(\alpha x)$ is positive and hence $K_{(2pi/\alpha)}(A\cosh(\alpha x))\in\mathbb{R}$. (ii) $K_{-(2pi/\alpha)}=K_{(2pi/\alpha)}$, so $W_0$ is even in $p$. (iii) the Wigner function is normalized,
\[
\iint W_0(x,p)\,\mathrm{d}x\,\mathrm{d}p=1.
\]
Substituting the explicit form of $W_0$ yields the identity
\[
\iint K_{(2pi/\alpha)}\!\bigl(A\cosh(\alpha x)\bigr)\,\mathrm{d}x\,\mathrm{d}p=\pi K_0(A).
\]

\begin{figure}[t]
    \centering
    \includegraphics[width=0.8\columnwidth]{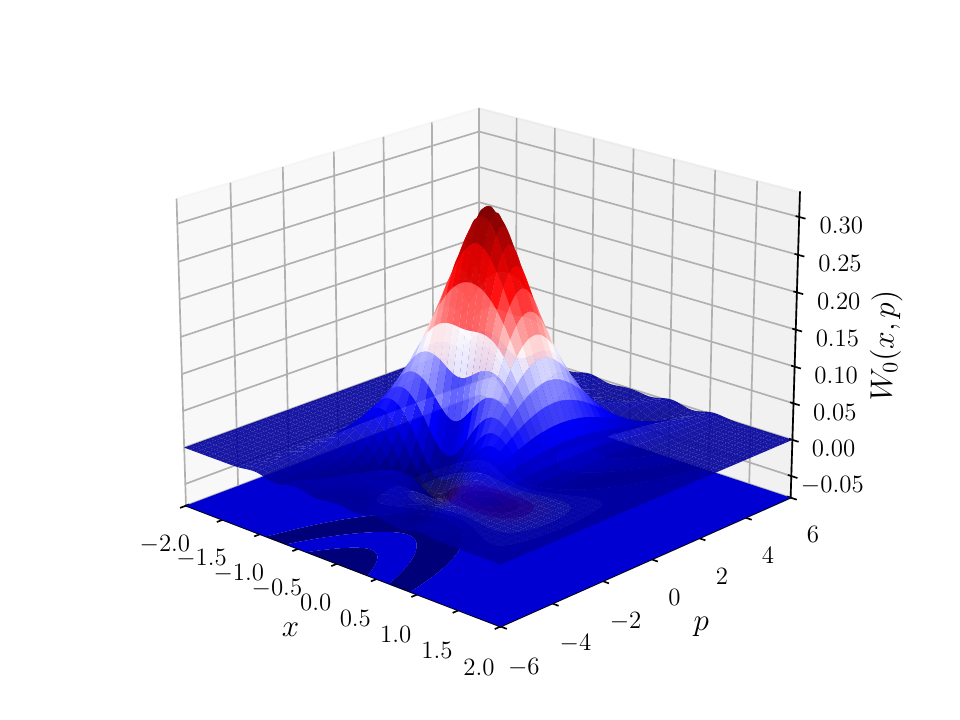}
    \caption{(Color online) Ground-state Wigner distribution $W_0(x,p)$ for $\alpha=5$ and $x_0=1$. Surface height equals $W_0(x,p)$; red (blue) shading indicates positive (negative) values. Axes are dimensionless and $W_0$ is normalized so that $\iint W_0\,\mathrm{d}x\,\mathrm{d}p=1$. The appearance of negative-valued regions provides a direct signature of nonclassicality for a pure state. The animations of the Wigner function is \cite{wignerAnimation}.}
    \label{fig5.2}
\end{figure}

Fig.~\ref{fig5.2} shows $W_0(x,p)$ for $\alpha=5$ and $x_0=1$. The Wigner function exhibits regions where $W_0<0$, a hallmark of nonclassicality for pure states. In a double-well potential, these negative regions arise from quantum interference between the two components of the ground-state wavefunction that delocalize across the central barrier. In phase space, this interference appears as oscillatory fringes—alternating positive and negative bands—most prominently near the barrier region (moderate $x$) and along the momentum direction; within each well, where the state locally resembles a Gaussian packet, $W_0$ is predominantly positive. The pattern is even in both $x$ and $p$ (since $\cosh \alpha x$ is even and $K_{-(2pi/\alpha)}=K_{(2pi/\alpha)}$), which underlies the quadrant reduction used to compute the integrated negativity:
\begin{equation}
    \nu \;=\; \frac{4}{\pi K_0(A)} \int_0^\infty\!\mathrm{d}\tilde{x} \int_0^\infty\!\mathrm{d}\tilde{p}\;
    \bigl|K_{i\tilde{p}}\!\bigl(A\cosh(2\tilde{x})\bigr)\bigr| \;-\; 1 .
    \label{eqn5.7}
\end{equation}

\begin{figure}[t]
    \centering
    \includegraphics[width=0.8\linewidth]{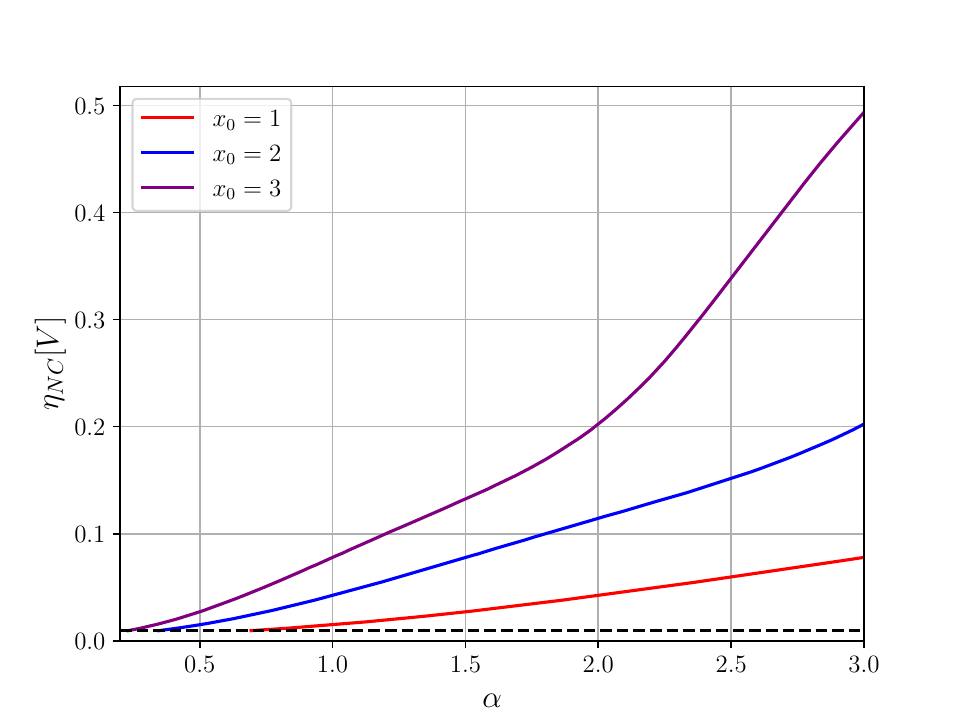}
    \caption{Nonclassicality measure $\eta_{\mathrm{NC}}$ as a function of $\alpha$ for the ground state with $x_0=1,2,3$. The dashed line marks $\eta_{\mathrm{NC}}\approx 0.009$.}
    \label{fig5.3}
\end{figure}

where $2\tilde{x} = \alpha x$ and $\tilde{p} = 2p/\alpha$. Using Eqs.~\eqref{eqn2.11} and \eqref{eqn5.7}, we evaluate the nonclassicality $\eta_{\mathrm{NC}}$. The integral in \eqref{eqn5.7} is evaluated numerically on finite windows $[0,x_\mathrm{max}]\times[0,p_\mathrm{max}]$ using Gauss-Legendre quadrature in both variables. These values were selected by monitoring the normalization condition
\begin{equation}
    \left|\int\int W_0(x,p)\mathrm{d}x\mathrm{d}p - 1\right|<\epsilon
\end{equation}

\begin{figure}
    \centering
    \includegraphics[width=0.8\columnwidth]{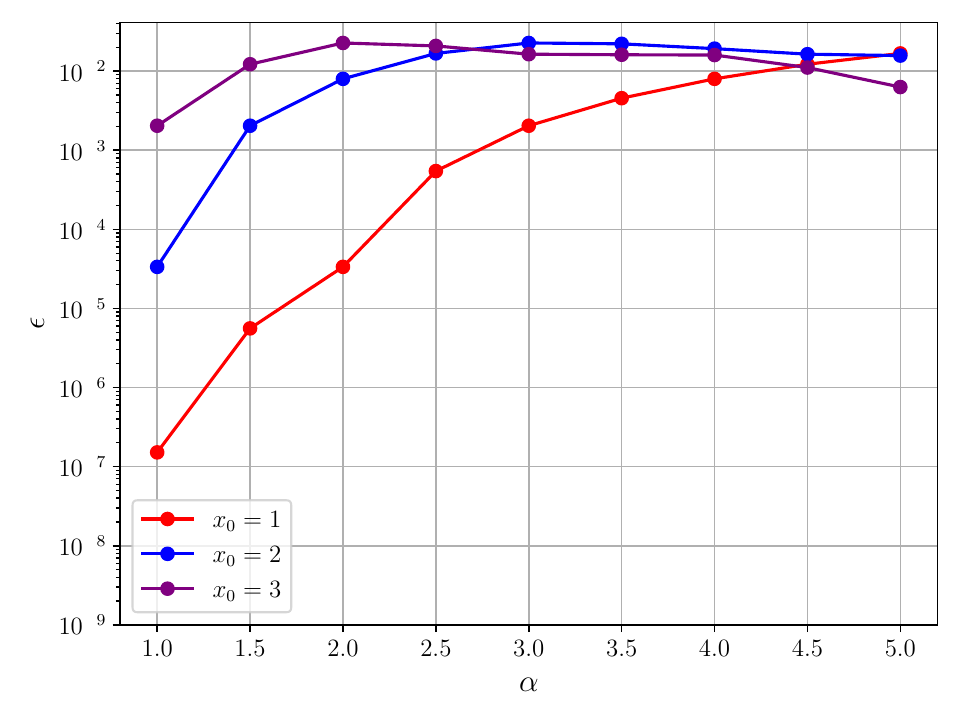}
    \caption{Log-scale error versus $\alpha$ for various $x_0$. The plots correspond to $x_\mathrm{max} = 15$ and $p_\mathrm{max} = 10$ with $N = 100$ divisions in both directions.}
    \label{wigErrFig}
\end{figure}

Fig. ~\ref{wigErrFig} display the plot of the log-scale error as a function of the parameter $\alpha$ for various $x_0$. The maximum integration limits in the $x$- and $p$-directions were chosen by ensuring sufficient decay of the Wigner distribution within the computational domain. As $\alpha$ and $x_0$ increase, the Wigner function exhibits more rapid oscillations, which leads to a corresponding increase in the numerical error.
This behavior is mainly associated with the limitations of adaptive quadrature for rapidly oscillatory integrals. 
In such cases, the nonuniform subdivisions generated by the adaptive algorithm may not adequately resolve the oscillatory structure of the integrand. A possible remedy is to employ uniformly spaced subdivisions, which can provide a more systematic sampling of the oscillations. However, this approach significantly increases the computational cost. Moreover, for the same number of integration points, uniformly spaced quadrature can be less efficient and may yield lower accuracy than adaptive quadrature in regions where the integrand varies smoothly. 
Therefore, adaptive quadrature was retained as a practical compromise between accuracy and computational efficiency.

As shown in Fig.~\ref{fig5.3}, $\eta_{\mathrm{NC}}$ rises monotonically with $\alpha$ for $x_0=1,2,3$, indicating that the ground state becomes increasingly nonclassical as the double well deepens and separates. Similar to the non-Gaussianity, the nonclassicality is an explicit function of $A(\alpha,x_0)$. In the shallow-well regime $\alpha \approx (\ln 2)/x_0$, all curves start from the same small baseline $\eta_{\mathrm{NC}}\!\approx\!0.009$ (dashed reference), consistent with a nearly Gaussian ground state with negligible Wigner negativity. Increasing $\alpha$ enhances interference between the two localized components, which increases the integrated negativity $\nu$ and the nonclassicality $\eta_{\mathrm{NC}}$.

\begin{figure}[t]
    \centering
    \includegraphics[width=0.8\linewidth]{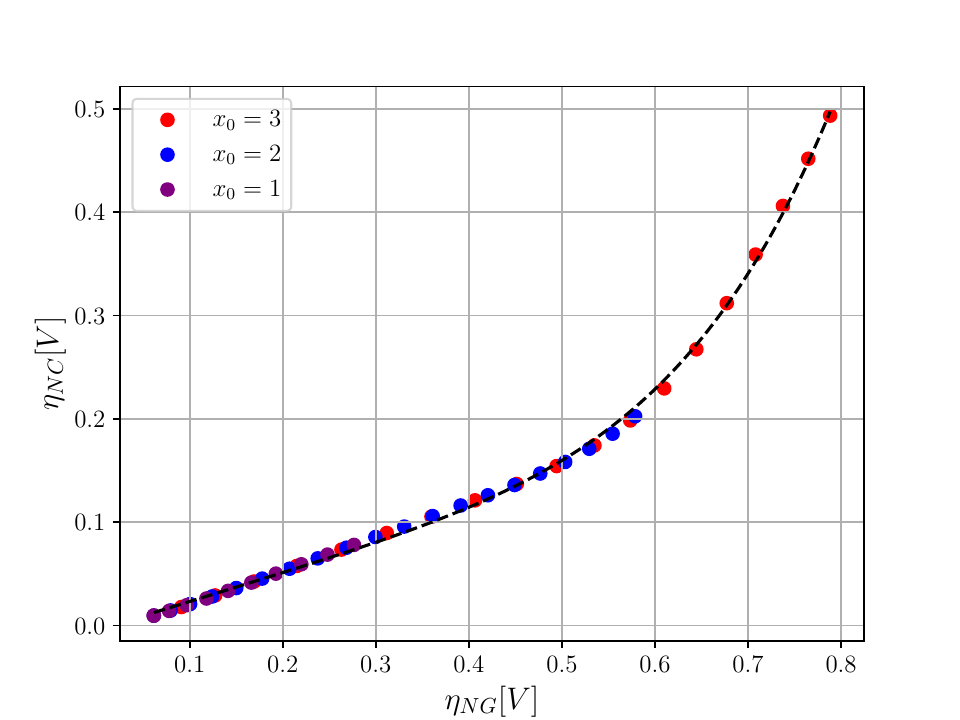}
    \caption{Parametric relation between non-Gaussianity $\eta_{\mathrm{NG}}$ and nonclassicality $\eta_{\mathrm{NC}}$. Symbols denote different $x_0$; the dashed curve is the fit to Eq.~\eqref{eqn5.8}.}
    \label{fig5.4}
\end{figure}

To quantify how non-Gaussianity and nonclassicality co-vary, we plot $\eta_{\mathrm{NC}}$ versus $\eta_{\mathrm{NG}}$ in Fig.~\ref{fig5.4}. Data for $x_0=1,2,3$ collapse onto a single monotone curve: the dependence is approximately linear for small $\eta_{\mathrm{NG}}$ and becomes superlinear at larger $\eta_{\mathrm{NG}}$. We model this behavior with
\begin{equation}
    \eta_{\mathrm{NC}} = a + b\,\eta_{\mathrm{NG}} + \eta_{\mathrm{NG}}^{\,c},
    \label{eqn5.8}
\end{equation}
which captures both the near-linear onset and the faster growth at higher non-Gaussianity. A least-squares fit yields

\begin{align*}
    a & = (-4.73 \times 10^{-3}) \pm (0.04 \times 10^{-3}) \\
    b & = 0.28 \pm (0.18 \times 10^{-5}) \\
    c & = 5.31 \pm (0.17 \times 10^{-3}).
\end{align*}

The small, slightly negative intercept \(a\) lies within systematic/numerical uncertainty and is consistent with the expectation that \(\eta_{\mathrm{NC}}\!\to\!0\) as \(\eta_{\mathrm{NG}}\!\to\!0\) (Gaussian limit).

The collapse of the three datasets onto a single parametric curve suggests that, once the state is characterized through resource measures rather than through the bare control variable $\alpha$, the relation between non-Gaussianity and Wigner negativity becomes largely insensitive to the geometric scale set by $x_0$ \emph{within the double-Morse family and the parameter range examined here} ($x_0=1,2,3$; $\alpha$ up to the values shown in Fig.~\ref{figNG}). We emphasize that Eq.~\eqref{eqn5.8} is an empirical fit to this specific model, motivated by the qualitatively similar linear-onset/superlinear-growth behavior noted for Duffing-like oscillators \cite{Teklu,Francesco}, and we do not have an independent theoretical derivation of the exponent $c$ or a proof that the functional form is universal. The value of $c$ obtained here should therefore be read as descriptive of the double-Morse ground state over the examined range, not as evidence of a general law for symmetric double-well potentials; extending the comparison to other double-well families (e.g., quartic or P\"oschl--Teller-type wells) would be needed before such a generalization could be supported.

\subsection{Entanglement potential}
\label{subsec:EPl}

In quantum optics, a single–mode state is considered nonclassical if, after passage through a passive linear–optical device, it can generate bipartite entanglement between two output modes. An operational quantifier of this resource is the \emph{entanglement potential} (EP): the entanglement created when the state interferes with the vacuum at a beam splitter 
 \cite{Aharonov1966,Kim2002,Asboth2005,Killoran2016PRL}.  It is sufficient and optimal to consider a lossless 50:50 beam splitter with vacuum in the auxiliary port; additional passive optics or extra vacuum ancillas cannot increase the bipartite entanglement across the output partition \cite{Asboth2005}. Related experimental and theoretical refinements appear in Refs.~\cite{ExpEP2024,RelNC2015}.

Let $|\psi\rangle$ be the input single–mode pure state and $|0\rangle$ the vacuum in the auxiliary input. The joint input is
\begin{equation}
|\Psi_{\text{in}}\rangle = |\psi\rangle \otimes |0\rangle .
\end{equation}
The beam-splitter unitary can be written as
\begin{equation}
\hat U(\xi)=\exp\!\big(\xi\,\hat a^\dagger \hat b-\xi^\ast\,\hat a \hat b^\dagger\big),\qquad \xi=\phi\,e^{i\theta},
\end{equation}
with $\hat a$ ($\hat b$) acting on output mode $A$ ($B$). Using Schwinger’s two–mode boson representation of $\mathrm{SU}(2)$, one obtains the disentangled form
\begin{equation}
\hat U(\xi)=\exp\!\big[-e^{-i\theta}\tan\phi\,\hat a\hat b^\dagger\big]\,
(\cos\phi)^{\hat a^\dagger\hat a-\hat b^\dagger\hat b}\,
\exp\!\big[e^{i\theta}\tan\phi\,\hat a^\dagger \hat b\big].
\end{equation}
A $50{:}50$ splitter corresponds to $\phi=\pi/4$; we denote $\hat B\equiv \hat U(\pi/4)$. The output state is then
\begin{equation}
|\Psi_{\text{out}}\rangle=\hat B\,|\Psi_{\text{in}}\rangle .
\end{equation}

For pure inputs, the EP reduces to the von Neumann entropy of either reduced state,
\begin{align}
E[\rho] &= -\mathrm{Tr}_B\!\left[\rho_B\log\rho_B\right]
= -\mathrm{Tr}_A\!\left[\rho_A\log\rho_A\right], \label{eq:EP-pure-entropy}
\end{align}
with $\rho=|\Psi_{\text{out}}\rangle\langle\Psi_{\text{out}}|$, $\rho_A=\mathrm{Tr}_B[\rho]$, and $\rho_B=\mathrm{Tr}_A[\rho]$. Equivalently,
\begin{equation}
E[\rho]=E\!\left[\hat B\left(|\psi\rangle\langle\psi|\ \otimes\ |0\rangle\langle 0|\right)\hat B^\dagger\right].
\end{equation}

\begin{figure}[t]
    \centering
    \includegraphics[width=0.8\linewidth]{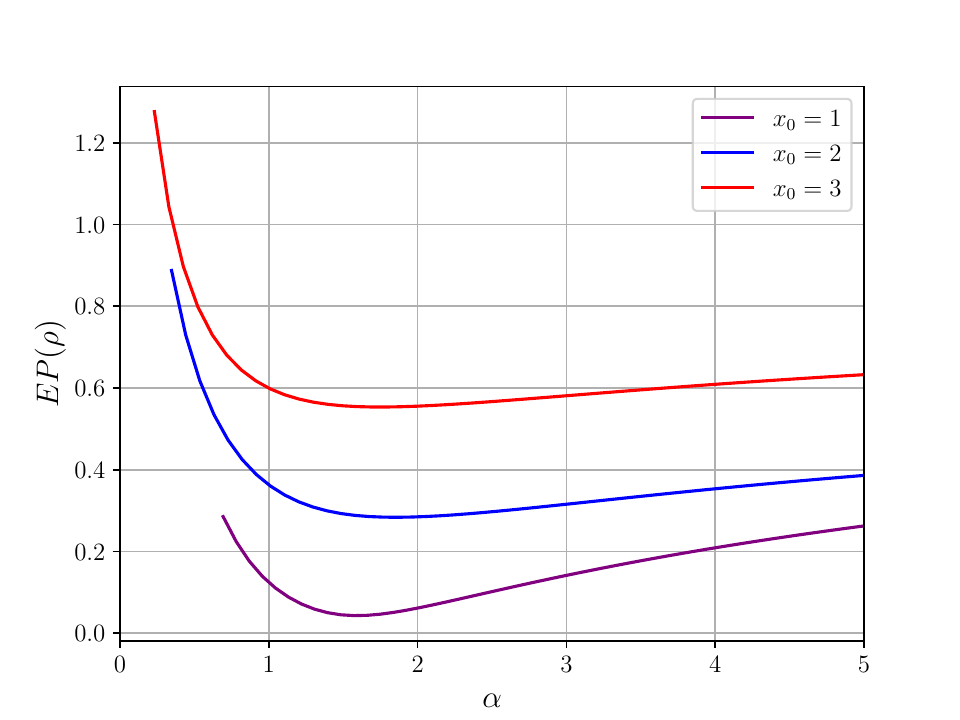}
   \caption{Entanglement potential $EP(\rho)$ generated by a $50{:}50$ beam splitter as a function of the state parameter $\alpha$. 
   Three curves, labeled by $x_0\in\{1,2,3\}$, show how this model parameter influences both the scale and the $\alpha$–dependence of the output entanglement.}
  \label{fig:EP_alpha}
\end{figure}

The numerical procedure of evaluating entanglement potential of a pure state is described as follows. The pure state is expanded in the Fock basis as
\begin{equation}
    |\psi \rangle = \sum_{n=0}^\infty c_n |n\rangle 
    \label{eqn_psi_series}
\end{equation}
where the series coefficients are $c_n = \langle n| \psi \rangle$. The Fock basis is the basis of the harmonic-oscillator associated with dimensionless quadratures satisfying the following 
\begin{equation}
    [\hat{q},\hat{p}] = i,\qquad \hat{a} = \frac{\hat{q}+i\hat{p}}{\sqrt{2}}, \qquad |n\rangle = \frac{(\hat{a}^\dag)^n}{\sqrt{n!}}|0\rangle  
\end{equation}
 Since the pure state is normalized, the normalization condition is
\begin{equation}
\sum_{n=0}^{\infty} |c_n|^2 = 1.
\end{equation}
For computation, the series in Eq. (\ref{eqn_psi_series}) is truncated to include finite terms $N_{\text{max}}$, which defines the finite dimension of the subspace. This approximation is carried by ensuring the norm of the pure state in the finite subspace is sufficiently close to unity. Mathematically, this condition is given by
\begin{equation}
    \left|\sum_{n=0}^{N_\text{max}}|c_n|^2 - 1\right| <\epsilon
    \label{eqn_control}
\end{equation}
where $\epsilon$ is the error tolerance. In the present calculation, a single fixed value of $N_{\text{max}}$ is not imposed for all parameter points. Instead, $N_{\text{max}}$ is increased adaptively for each pair $(\alpha,x_0)$ until Eq.~\eqref{eqn_control} is satisfied. The entanglement-potential values in Fig.~\ref{fig:EP_alpha} were retained only after this discarded-norm criterion was satisfied. This norm-based criterion was used as the truncation-convergence check for the von Neumann entropy of the beam-splitter output.

Fig.~\ref{fig:EP_alpha} shows the entanglement potential $EP(\rho)$ as a function of the width parameter $\alpha$ for three values of $x_0$. The curves 
display a shallow nonmonotonic feature at small $\alpha$, where $EP(\rho)$ first decreases before increasing gradually with $\alpha$. This behavior reflects a 
crossover in the ground-state wavefunction as the system changes from a nearly single-well configuration to a more pronounced double-well structure. Once this
crossover is passed, the increasing separation of the wells and the coherence between the localized components of the ground state lead to enhanced output 
entanglement at the beam splitter. Larger values of $x_0$ produce larger entanglement potentials, with the ordering $x_0=3>x_0=2>x_0=1$ preserved over
the plotted range. Thus, beyond the low-$\alpha$ crossover, increasing $\alpha$ enhances the two-mode entanglement generated from the double-Morse ground 
state, complementing the trends observed for non-Gaussianity and nonclassicality.

\section{Local quantum estimation theory}
\label{sec3}
Quantum technologies rely on properties of individual quantum systems and can 
often outperform any conventional technology in precision measurements and 
information processing. Relevant examples may be found in quantum metrology, 
which allows one to achieve much more sensitive parameter estimation than any 
conventional techniques. Indeed, quantum properties of physical systems are 
{\em fragile} to decoherence, which in turn makes suitably engineered sensors 
exceptionally sensitive to small parameter changes. This general viewpoint remains 
relevant in noisy environments as well, where non-Markovian memory effects may 
restore a genuine quantum enhancement in parameter estimation \cite{Chin2012}.

Local quantum estimation theory (QET) looks for the measurement maximizing the Fisher information, an 
abstract quantity that measures the maximum information about the parameter we 
wish to estimate from a given measurement procedure, thus minimizing the variance 
of the estimator. Roughly speaking, one may expect local QET to provide better 
performances since the optimization concerns a specific value of the parameter, 
with some adaptive or feedback mechanism assuring the achievability of the ultimate bound. 

Local QET has been successfully applied to estimation problems in open
quantum systems, non-unitary processes and nonlinear quantities as
entanglement~\cite{monras,Paris2009,dorner,brida,Brivio,carmen11,genoni11,tsang,genoni12,genoni13,Pinel,Candeloro,asjad2023joint,Simone25,Chabane} and for a
closed system, evolving under a unitary transformation~\cite{Boxio}.
The geometric structure of QET has been exploited to assess quantum
criticality as a resource for quantum estimation~\cite{ZP08}.

\subsection{Quantum Fisher information}

In classical estimation theory, a family of distributions $p(x|\theta)$ has classical Fisher
information (CFI)
\begin{align}
F_\theta
&= \int p(x|\theta)\left(\frac{\partial}{\partial\theta}\log p(x|\theta)\right)^{2}dx\\
&= \int \frac{1}{p(x|\theta)}\left(\frac{\partial p(x|\theta)}{\partial\theta}\right)^{2}dx,
\label{eq:cfi}
\end{align}
and any unbiased estimator $\hat{\theta}$ obeys the Cram\'{e}r-Rao bound 
\begin{equation}
\mathrm{Var}(\hat{\theta}) \ge \frac{1}{F_\theta}.
\label{eq:crb}
\end{equation}
%%%%%
In the quantum setting, the parameter $\theta$ is encoded in a state $\rho_{\theta}$. The symmetric logarithmic derivative (SLD) $L_\theta$ is defined implicitly by ~\cite{Holevo2011,Paris2009}
\begin{equation}
\partial_\theta \rho_\theta = \tfrac{1}{2}\big(L_\theta \rho_\theta + \rho_\theta L_\theta\big).
\label{eq:sld}
\end{equation} 
The quantum Fisher information (QFI) is
\begin{equation}
\mathcal{F}(\theta) = \mathrm{Tr}\!\left[\rho_\theta L_\theta^2\right],
\label{eq:qfi}
\end{equation}
and the quantum Cramér--Rao bound (QCRB) implies that for any (locally) unbiased estimator $\hat\theta$,
\begin{equation}
\mathrm{Var}(\hat\theta) \ge \frac{1}{\mathcal{F}(\theta)}.
\label{eq:qcrb}
\end{equation}

By writing the quantum state with the spectral decomposition, $\rho_\theta=\sum_{n}p_{n}|\psi_{n}\rangle\langle\psi_{n}|$, 
a widely used expression for the QFI is
\begin{align}
F(\rho_\theta)
&= \sum_{n} \frac{(\partial_\theta p_n)^2}{p_n}
+ \sum_{n} 4 p_n \,\langle \partial_\theta \psi_n \mid \partial_\theta \psi_n \rangle\\
&- \sum_{m,n} \frac{8 p_m p_n}{p_m + p_n} \, \bigl|\langle \partial_\theta \psi_m \mid \psi_n \rangle\bigr|^2 .
    \label{eqn14}
\end{align}
In the case of a pure state $|\psi_{\theta}\rangle$, the above expression further simplifies to~\cite{Paris2009} 

\begin{equation}
\mathcal{F}_\theta
= 4\left(
\langle \partial_\theta \psi_\theta \mid \partial_\theta \psi_\theta \rangle
- \bigl\lvert \langle \psi_\theta \mid \partial_\theta \psi_\theta \rangle \bigr\rvert^{2}
\right).
\label{eqn15}
\end{equation}
Furthermore, note that the real part of $\langle \partial_\theta \psi_\theta \mid \psi_\theta \rangle$ is zero for real states, hence the 
QFI is just given by
\begin{equation}
 \mathcal{F}_\theta=4\langle \partial_{\theta}\psi_\theta \mid \partial_\theta \psi_\theta \rangle.
\label{eqn16}
\end{equation}

In integral form, QFI is evaluated as 
\begin{equation}
    \mathcal{F}(\alpha) = 4\int_{-\infty}^{+\infty} \left(\frac{\partial \psi_0(x)}{\partial \alpha}\right)^2\ dx
    \label{qfiGen} 
\end{equation}
where the integrand takes the form 
\begin{equation}
    \frac{\partial \psi_0(x)}{\partial \alpha} = \frac{\psi_0(x)}{2\alpha}\left[1-A\alpha x_0\frac{K_1(A)}{K_0(A)} + A\alpha x_0 \cosh\alpha x - A\alpha x\sinh\alpha x\right]
    \label{qfiIntegrand}
\end{equation}
The QFI is evaluated using adaptive quadrature techniques described earlier.

For the family of real ground states $\psi_\alpha(x)$, the classical Fisher information associated with position measurements,
\begin{equation}
F_x(\alpha)=\int dx\,p(x|\alpha)\left[\partial_\alpha\ln p(x|\alpha)\right]^2\,
,p(x|\alpha)=|\psi_\alpha(x)|^2,
\label{eq:cfi-position}
\end{equation}
coincides with the QFI. Indeed, since $\psi_\alpha(x)$ is real and normalized, one has
\begin{equation}
\langle \psi_\alpha|\partial_\alpha\psi_\alpha\rangle=0,
\label{eq:orthogonality-alpha}
\end{equation}
and therefore
\begin{equation}
F_x(\alpha)=4\int dx\,\bigl(\partial_\alpha\psi_\alpha(x)\bigr)^2=\mathcal{F}(\alpha).
\label{eq:Fx-equals-QFI}
\end{equation}
This shows that position measurements are optimal and saturate the quantum Cram\'er--Rao bound.

\begin{figure}
    \centering
    \includegraphics[width=0.8\columnwidth]{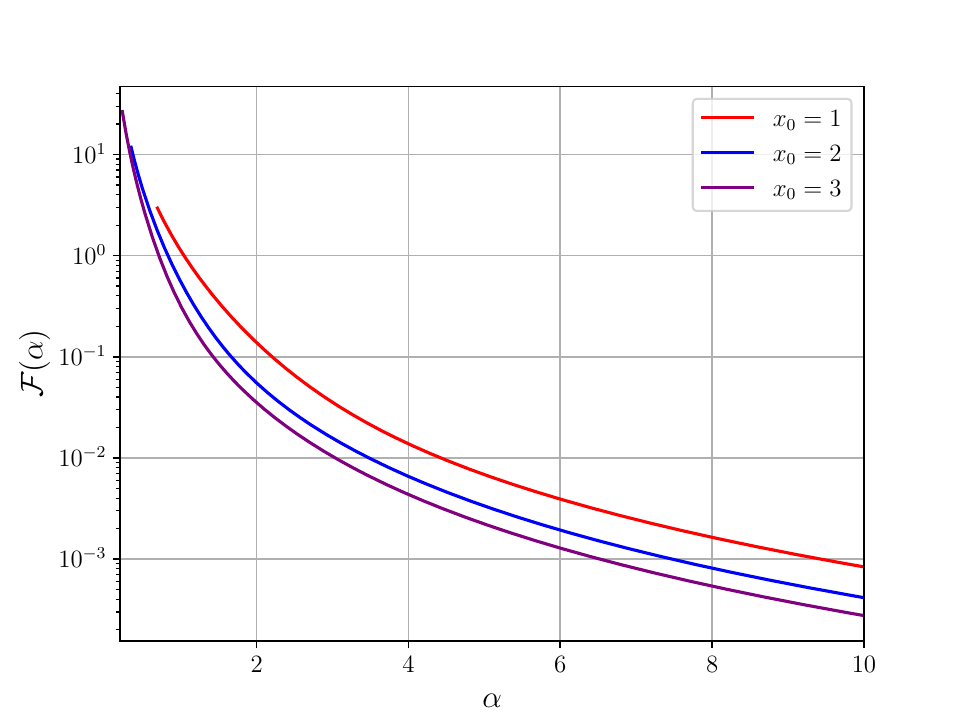}
    \caption{Fisher information of the ground-state wavefunction as a function of $\alpha$ for different values of $x_0$.}
    \label{fig6}
\end{figure}

As shown in Fig.~\ref{fig6}, the QFI of the ground state, denoted by $\mathcal{F}(\alpha)$, decreases with $\alpha$ for all well separations $x_0\in\{1,2,3\}$. This monotonic decrease reflects a diminishing sensitivity of the state to changes in $\alpha$ and, via the quantum Cram\'er--Rao bound, implies that the minimum achievable variance of any locally unbiased estimator increases as $\alpha$ grows, i.e., $\mathrm{Var}(\hat{\alpha})\ge [\mathcal{F}(\alpha)]^{-1}$. Conversely, in the shallow-well regime (small $\alpha$), the QFI is larger, yielding tighter lower bounds on the estimation error. In view of Eq.~\eqref{eq:Fx-equals-QFI}, the optimal measurement is a position measurement.

To probe the deep-well regime, it is natural to consider instead the dimensionless control parameter
\begin{equation}
A=2e^{-\alpha x_0},
\label{eq:Adef_met}
\end{equation}
which enters directly into the ground-state wavefunction and the phase-space expressions. For fixed and independently calibrated $x_0$, the map $\alpha\mapsto A$ is smooth, monotone, and one-to-one throughout the double-well domain. Since Fisher information depends on the chosen parameterization, a change of variables $\beta=f(\alpha)$ transforms both the classical FI and the QFI according to the chain rule,
\begin{equation}
F_{\beta}=F_{\alpha}\left(\frac{d\alpha}{d\beta}\right)^2,
\qquad
\mathcal{F}(\beta)=\mathcal{F}(\alpha)\left(\frac{d\alpha}{d\beta}\right)^2.
\label{eq:chainrule_met}
\end{equation}
With $\beta\equiv A$ from Eq.~\eqref{eq:Adef_met}, one has
\begin{equation}
\frac{dA}{d\alpha}=-x_0A,
\qquad
\frac{d\alpha}{dA}=-\frac{1}{x_0A},
\label{eq:Aalpha-derivatives}
\end{equation}
and therefore
\begin{equation}
\mathcal{F}(A)=\mathcal{F}(\alpha)\left(\frac{d\alpha}{dA}\right)^2=\frac{\mathcal{F}(\alpha)}{x_0^2A^2}.
\label{eq:F_A_from_F_alpha}
\end{equation}
Because $A$ decays exponentially with $\alpha$, the prefactor $1/A^2$ grows rapidly as $\alpha$ increases. As a consequence, the QFI associated with estimating $A$ may increase with $\alpha$ even when $\mathcal{F}(\alpha)$ decreases. This does not contradict the coordinate invariance of statistical distinguishability; rather, it reflects that $A$ is a compressed nonlinear function of $\alpha$.

\begin{figure}[htbp]
\centering
\includegraphics[width=0.92\linewidth]{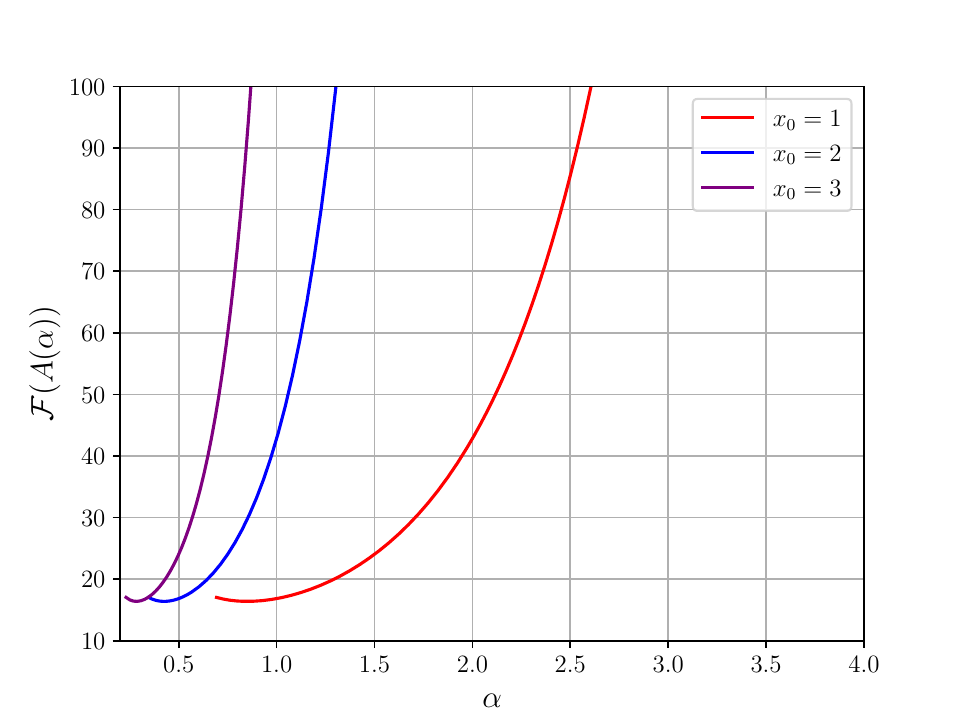}
\caption{Quantum Fisher information for estimating the composite control parameter $A=2e^{-\alpha x_0}$, shown as a function of $\alpha$ for representative values of $x_0$.}
\label{fig:FI_A_vs_alpha}
\end{figure}

\begin{figure}[htbp]
\centering
\includegraphics[width=0.92\linewidth]{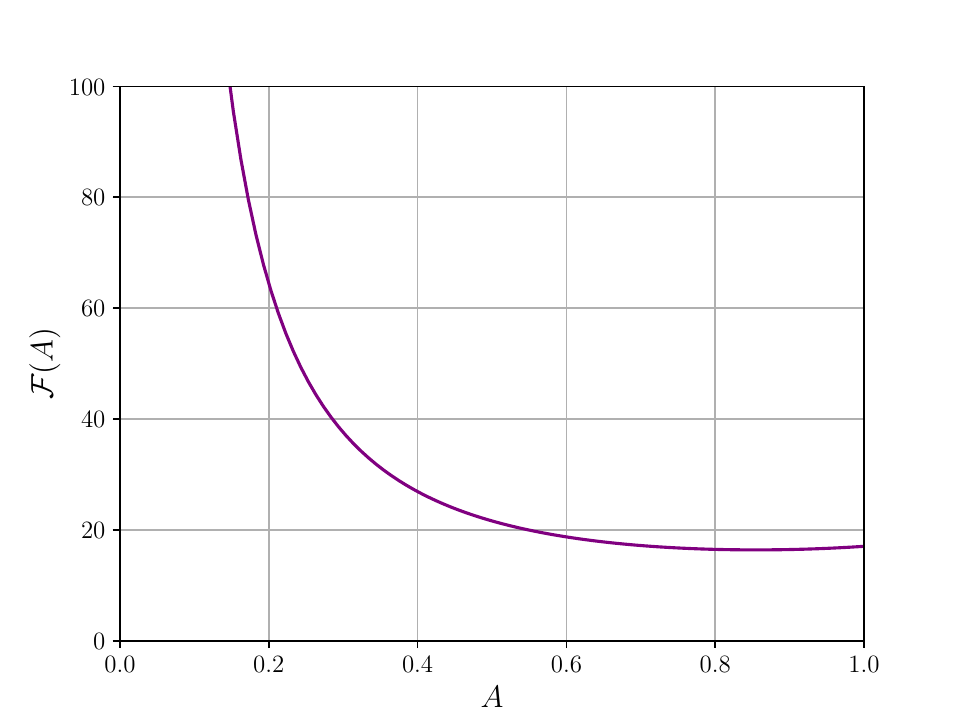}
\caption{Quantum Fisher information for estimating $A$, shown directly as a function of $A$.}
\label{fig:FI_A_vs_A}
\end{figure}

Fig.~\ref{fig:FI_A_vs_alpha} shows the QFI for estimating $A$ as a function of $\alpha$ for representative values of $x_0$, while Fig.~\ref{fig:FI_A_vs_A} shows the same quantity directly as a function of $A$. These two representations are equivalent because $A=A(\alpha)$ is monotone in $\alpha$ for fixed $x_0$.

To estimate $\alpha$ from a measurement of $A$, one uses
\begin{equation}
\alpha=-\frac{1}{x_0}\ln\!\left(\frac{A}{2}\right).
\label{eq:alpha_from_A_met}
\end{equation}
Hence, estimating $A$ determines $\alpha$ only if $x_0$ is known. In many engineered platforms, $x_0$ is a geometric or trap-design parameter that can be calibrated independently, for example by static potential mapping or direct imaging of the well minima. In that common setting, single-parameter estimation of $A$ is well posed and can yield an operational metrological advantage in the deep double-well regime. If one wishes instead to report the precision on $\alpha$, the Cram\'er--Rao bound is unchanged by reparameterization. For $N$ independent repetitions,
\begin{equation}
\mathrm{Var}(\hat{\alpha})\ge \frac{1}{N\mathcal{F}(\alpha)}=\left(\frac{d\alpha}{dA}\right)^2\frac{1}{N\mathcal{F}(A)}.
\label{eq:crb-reparam-alpha}
\end{equation}
Thus, the increase of $\mathcal{F}(A)$ at small $A$ should be interpreted as enhanced sensitivity to the experimentally relevant control knob $A$, rather than as an improvement in the precision bound for $\alpha$ itself. If the final reported parameter is $\alpha$, the uncertainty must be propagated through Eq. ~(\ref{eq:alpha_from_A_met}).

If $x_0$ is not known a priori, then $\alpha$ is no longer a single-parameter quantity inferred from $A$ alone. In that case, one must either calibrate $x_0$ independently or treat the problem as a two-parameter estimation task. At first order, error propagation from Eq.~\eqref{eq:alpha_from_A_met} gives

\begin{equation}
\begin{aligned}
\mathrm{Var}(\hat{\alpha})
&\approx
\left(\frac{\partial\alpha}{\partial A}\right)^2\mathrm{Var}(\hat{A})+
\left(\frac{\partial\alpha}{\partial x_0}\right)^2\mathrm{Var}(\hat{x}_0)\\
&\quad
+2\frac{\partial\alpha}{\partial A}\frac{\partial\alpha}{\partial x_0}\,\mathrm{Cov}(\hat{A},\hat{x}_0)\\
&=
\frac{\mathrm{Var}(\hat{A})}{x_0^2A^2}+
\frac{\ln^2(A/2)}{x_0^4}\,\mathrm{Var}(\hat{x}_0)\\
&\quad
+2\frac{\ln(A/2)}{x_0^3A}\,\mathrm{Cov}(\hat{A},\hat{x}_0)
\end{aligned}
\label{eq:errorprop_met}
\end{equation}
where $\partial\alpha/\partial A=-1/(x_0A)$ and $\partial\alpha/\partial x_0=\ln(A/2)/x_0^2$. If $\hat{A}$ and $\hat{x}_0$ are obtained independently, the covariance term vanishes. Equation~\eqref{eq:errorprop_met} makes the engineering message explicit: in the regime $A\to 0$ (large $\alpha$), the inferred value of $\alpha$ can become highly sensitive to even small uncertainty in $x_0$ because of the logarithmic factor.

A compact way to quantify the effect of an unknown $x_0$ is through the Fisher-information matrix. For $\bm{\theta}=(A,x_0)$, define the classical FI matrix $\mathbf{F}$ with elements
\begin{equation}
F_{ij}=\int dx\ p(x|\bm{\theta})\
\frac{\partial\ln p(x|\bm{\theta})}{\partial\theta_i}\
\frac{\partial\ln p(x|\bm{\theta})}{\partial\theta_j}.
\label{eq:FIM-2param}
\end{equation}
For $N$ repetitions, the matrix Cram\'er--Rao bound reads $\mathrm{Cov}(\hat{\bm{\theta}})\succeq [N\mathbf{F}]^{-1}$. If $A$ is the parameter of interest and $x_0$ is a nuisance parameter, the effective information is reduced according to the Schur complement, giving
\begin{equation}
\mathrm{Var}(\hat{A})\ge \frac{1}{N\left(F_{AA}-\dfrac{F_{Ax_0}^2}{F_{x_0x_0}}\right)}.
\label{eq:schur_met}
\end{equation}
An analogous expression holds if $\alpha$ is the parameter of interest and $x_0$ is treated as a nuisance parameter. A full quantum treatment may be obtained by replacing $\mathbf{F}$ with the quantum Fisher-information matrix and applying the corresponding matrix quantum Cram\'er--Rao bound.

\subsection{A concrete sensing interpretation}
\label{subsec:sensing}

The metrological results acquire a direct physical meaning when the double-Morse potential is viewed as an effective sensor Hamiltonian. In proton-transfer or ferroelectric realizations, for example, $\alpha$ may be interpreted as an inverse barrier width controlled by strain, bias field, or local electrostatic environment, while $A=2e^{-\alpha x_0}$ summarizes the experimentally adjusted barrier profile. The quantities plotted in Figs.~\ref{fig6}--\ref{fig:FI_A_vs_A} then translate directly into sensitivity bounds for estimating that control knob.

For $N$ independent preparations and position measurements, the quantum Cram\'er--Rao bound gives
\begin{equation}
\delta\alpha_{\min}=\frac{1}{\sqrt{N\mathcal{F}(\alpha)}},
\qquad
\delta A_{\min}=\frac{1}{\sqrt{N\mathcal{F}(A)}}.
\label{eq:sensing_bounds}
\end{equation}
Hence a straightforward increase from $N=10^3$ to $N=10^4$ reduces the error bar by the expected factor $\sqrt{10}$. More importantly, the relevant parameter depends on the experimental question: shallow double wells are optimal if the aim is to estimate $\alpha$ itself, whereas deep double wells are advantageous when the directly tunable quantity is the barrier-control parameter $A$. This distinction is operational rather than semantic, because in many devices the laboratory knob is a voltage, laser intensity, or strain bias that primarily modifies the barrier profile instead of a directly measured coordinate.

For comparison with a harmonic reference sensor, it is natural to define a relative metrological gain
\begin{equation}
G_{\mathrm{dB}} = 10\log_{10}\!\left[\frac{\mathcal{F}_{\mathrm{DM}}}{\mathcal{F}_{\mathrm{ref}}}\right],
\label{eq:gain_db}
\end{equation}
where $\mathcal{F}_{\mathrm{ref}}$ denotes the Fisher information of the chosen harmonic benchmark under the same readout protocol. A platform-specific evaluation of $G_{\mathrm{dB}}$ requires fixing the calibration convention for the reference oscillator, but the present results already show where the gain should occur: namely in the regime where the double-well structure produces simultaneously large non-Gaussianity, Wigner negativity, and enhanced sensitivity to the experimentally accessible control parameter $A$. This makes the double-Morse family a practically actionable candidate for barrier-based quantum sensing.

\subsection{Experimental realizations and physical scales}
\label{subsec:exp_realizations}

Beyond its molecular origin, the double-Morse profile may be engineered as an effective one-dimensional potential in several bosonic platforms of current interest. In ultracold-atom settings, painted potentials, spatial light modulators, and superlattice geometries can produce two minima separated by a controllable barrier, making $\alpha$ a natural inverse-length parameter for the barrier width while $x_0$ fixes the geometric half-separation. In trapped-ion architectures, segmented electrodes and tailored axial confinement allow one to generate anharmonic motional potentials with an adjustable central barrier \cite{RevModPhys.75.281,Home_2011}. Related ideas also arise in optomechanical and electromechanical devices, where static biasing and radiation-pressure-induced softening may produce bistable effective potentials for a mechanical mode \cite{RevModPhys.86.1391,Rogers2014,Sankey2010}. Finally, the same parametrization is well suited to proton-transfer and ferroelectric tunneling models, in which the barrier profile is directly tied to bond geometry and local fields \cite{matsushita1982note,robertson1981analysis,goryainov2012model}. These examples broaden the relevance of the present analysis beyond the molecular setting and connect it to quantum sensing and continuous-variable state engineering.

For experimental translation, it is convenient to interpret $\alpha$ as an inverse length, $x_0$ as the geometric half-separation between the wells, and $D/\hbar\omega$ as a dimensionless depth once a local oscillator scale $\omega$ has been identified. In the bistable regime, the barrier height is $V_{\mathrm{DM}}(0)=D(1-A)^2$ and the effective control parameter $A=2e^{-\alpha x_0}$ compactly captures how the barrier responds to a tunable voltage, laser intensity, strain field, or trap setting. This parametrization is particularly useful because the exact ground-state expressions derived below can be mapped onto different hardware platforms with no change in the analytic structure.

Compared with Duffing or quartic models, the double-Morse oscillator offers the attractive combination of natural bistability and exact solvability at the level relevant for the present resource analysis. This makes it a useful reference model for experimental groups seeking non-Gaussian-state generators or tunable double-well sensors.

\section{Conclusion}
\label{sec4}
In this work, we have shown that the double-Morse potential provides a practical and controllable platform for creating non-Gaussian quantum states. The Wigner-function negativities show that when the asymmetry (width parameter) $\alpha$ is tuned, the ground state significantly deviates from Gaussianity and acquires prominent nonclassical properties. Quantitative measures of non-Gaussianity and nonclassicality exhibit a monotonic relationship with each other over the examined parameter regimes, establishing an operational connection between these two resources in this model.

The same control parameter also governs the state's capacity to generate bipartite entanglement at a passive $50:50$ beam splitter: the entanglement potential $EP(\rho)$ exhibits a shallow low-$\alpha$ dip, followed by a steady increase and a tendency toward saturation. Larger $x_0$ generally yields higher output entanglement (Fig.~\ref{fig:EP_alpha}). Thus, beyond the small-$\alpha$ crossover region, increased structural asymmetry enhances the accessible nonclassical resources.\\

To characterize the metrological performance, we studied the estimation of the structural parameter $\alpha$ from the ground state. We found that position measurements are optimal in this model and saturate the Cram\'er--Rao bound. The corresponding quantum Fisher information $\mathcal{F}(\alpha)$ decreases with $\alpha$, which means that direct estimation of $\alpha$ is most precise in the shallow-well regime. For deep double wells, it is more natural to estimate the experimentally relevant control parameter $A=2e^{-\alpha x_0}$, whose Fisher information can increase with $\alpha$ because of the nonlinear reparameterization. Therefore, the metrological advantage of the double-Morse platform depends on whether the target parameter is $\alpha$ itself or the derived control knob $A$, with the latter requiring independent calibration of $x_0$.
\\
Our findings also point to a clear experimental roadmap. The double-Morse profile can be viewed as an effective model for engineered double wells in ultracold atoms, trapped ions, optomechanical/electromechanical devices, and proton-transfer systems, with $\alpha$ and $A$ playing the role of directly tunable barrier-control parameters. In this sense, the model complements more familiar Duffing-type descriptions by combining natural bistability with exact solvability. The sensing interpretation clarifies when the relevant target parameter is $\alpha$ and when it is instead the experimentally accessible control knob $A$. Taken together, these features make the double-Morse oscillator a viable and tunable source of non-Gaussianity for precision metrology \cite{Chabane,Volkoff_2025} and quantum simulation, while the resulting dynamics offer insight into matter-wave packet control and potential applications in quantum information processing and computing \cite{li2013quantum,Averbukh1989,Robinett2004,Stapelfeldt2003,Koch2019,DeMille2002,BerryKlein1996}.

\subsection{Outlook: open-system extension}
\label{subsec:open_system}
A natural direction beyond the present closed-system analysis is to quantify how amplitude damping and dephasing degrade the resource measures studied above. A full open-system study requires numerical time evolution of the density matrix and lies outside the scope of the present manuscript; we outline it here only as a possible extension, not as an additional result. A minimal Markovian model may be written as
\begin{equation}
\dot{\rho}=-\frac{i}{\hbar}[\hat{H}_{\mathrm{DM}},\rho]
+\gamma(\bar{n}+1)\mathcal{L}[\hat{a}]\rho
+\gamma\bar{n}\,\mathcal{L}[\hat{a}^{\dagger}]\rho
+\kappa_{\phi}\,\mathcal{L}[\hat{a}^{\dagger}\hat{a}]\rho,
\label{eq:lindblad_DM}
\end{equation}
where
\begin{equation}
    \mathcal{L}[\hat O]\rho=\hat O\rho\hat O^\dagger-
    \frac{1}{2}\{\hat O^\dagger\hat O,\rho\}.
\end{equation}
We use $\mathcal{L}$ rather than $\mathcal{D}$ for the dissipator to avoid confusion with the potential depth $D$. The operator $\hat a$ is a reference-mode annihilation operator, and its use is appropriate only when the bath couples weakly to the chosen harmonic reference mode. Since the double-Morse ground state is delocalized across both wells, a more faithful model may instead require a global normal-mode basis or a localized two-mode description with well-to-well dissipative coupling. Consequently, no quantitative conclusion about robustness under damping or dephasing is drawn from Eq.~\eqref{eq:lindblad_DM}; establishing the decay of $\eta_{\mathrm{NG}}$, $\eta_{\mathrm{NC}}$, and $EP(\rho)$ under this or a more faithful dissipative model, together with a justified choice of reference mode, is left for future work.

\begin{acknowledgments}
\noindent\textbf{Funding:}  This research was funded by Khalifa University of Science and Technology through the Project ID: KU-INT-RIG-2024-8474000739.
\end{acknowledgments}

\appendix
\section{Covariance Matrix}
\label{covar_matrix}

Non-Gaussianity of the quantum state $\psi(x)$ is measured by considering a reference Gaussian state that has the same mean vector and covariance matrix as $\psi(x)$. The covariance matrix is
\begin{equation*}
\sigma =
\begin{pmatrix}
\langle \hat{x}^2 \rangle - \langle \hat{x} \rangle^2 & \frac{1}{2}\langle \{\hat{x},\hat{p}\} \rangle  - \langle \hat{x} \rangle \langle \hat{p} \rangle \\
\frac{1}{2}\langle \{\hat{p},\hat{x}\} \rangle  - \langle \hat{p} \rangle \langle \hat{x} \rangle & \langle \hat{p}^2 \rangle - \langle \hat{p} \rangle^2
\end{pmatrix}.
\end{equation*}

The operators in the position representation are
\begin{align*}
    \hat{x} & = x, \\
    \hat{p} & = -i\,\frac{\partial}{\partial x}.
\end{align*}
Using $\langle \hat{O} \rangle = \langle \psi_0 | \hat{O} | \psi_0 \rangle = \int_{-\infty}^{+\infty} \psi_0^\ast(x) \hat{O}_x \psi_0(x)\ dx$, the following expressions are obtained:
\begin{align}
    \langle \hat{x} \rangle & = \frac{\alpha}{2K_0(A)}\int_{-\infty}^{+\infty} e^{-A\cosh \alpha x}\, x\, dx = 0, \label{eqna1} \\
    \langle \hat{p} \rangle & = \frac{-A\alpha^2i}{2K_0(A)}\int_{-\infty}^{+\infty} e^{-A\cosh \alpha x}\, \sinh \alpha x\, dx = 0, \label{eqna2}\\
    \langle \hat{x}^2 \rangle & = \frac{\alpha}{2K_0(A)}\int_{-\infty}^{+\infty} x^2 e^{-A\cosh \alpha x}\, dx \nonumber \\
    & = \frac{1}{\alpha^2 K_0(A)} \int_0^{\infty} y^2 e^{-A\cosh y}\, dy, \label{eqna3} \\
    \langle \hat{p}^2 \rangle & = \frac{A\alpha^3}{8K_0(A)}\int_{-\infty}^{+\infty} e^{-A\cosh \alpha x}\!\left(2\cosh \alpha x - A \sinh^2 \alpha x\right)\! dx \nonumber \\
    & =  \frac{\alpha^2}{4K_0(A)} \left(2AK_1(A) - \frac{A^2}{2}(K_2(A)-K_0(A))\right). \label{eqna4}
\end{align}

Equations~\eqref{eqna1} and \eqref{eqna2} vanish because their integrands are odd. The expression for $\langle \hat{p}^2 \rangle$ follows from Eq.~\eqref{eqn:handbook}. Using the canonical commutation relation $[\hat{x},\hat{p}]=i$, the anti-commutator simplifies to $\{\hat{x},\hat{p}\}=2\hat{x}\hat{p}-i$. Its expectation value is
\begin{align}
    \langle \{\hat{x},\hat{p}\} \rangle & = \langle 2\hat{x}\hat{p} - i \rangle \nonumber\\
    & = 2\langle \hat{x}\hat{p} \rangle - i \nonumber \\
    & = \frac{A\alpha^2i}{K_0(A)} \int_{-\infty}^{+\infty} e^{-A\cosh \alpha x}\, x \, \sinh \alpha x\, dx - i \nonumber \\
    & = \frac{Ai}{K_0(A)} \int_0^{\infty} e^{-A\cosh y}\, y\, \sinh y\, dy - i. \label{eqna5}
\end{align}
The expressions in Eqs.~\eqref{eqna3} and \eqref{eqna5} follow from the substitution $y=\alpha x$. Although these integrals may be evaluated numerically (e.g., by Gauss--Legendre quadrature), it is preferable to obtain analytic forms using Feynman's trick (differentiation under the integral sign). Consider
\begin{equation}
    I(b) = \int_0^\infty e^{-A\cosh (b x)}\, dx.
    \label{eqna6}
\end{equation}
Differentiating with respect to $b$ gives
\begin{equation}
    \frac{dI}{db} = -A\int_0^\infty e^{-A\cosh(bx)} \, \sinh (b x)\, x\, dx.
    \label{eqna7}
\end{equation}
Setting $b=1$ yields
\begin{equation}
    \int_0^\infty e^{-A\cosh x}\, x \sinh x\, dx = -\frac{1}{A} \left.\frac{dI}{db}\right|_{b = 1}.
    \label{eqna8}
\end{equation}
Using Eq.~\eqref{eqn:handbook} to evaluate $I(b)$, one finds
\begin{equation}
    \int_0^\infty e^{-A\cosh x}\, x \sinh x\, dx = \frac{K_0(A)}{A}.
    \label{eqna9}
\end{equation}
Substituting Eq.~\eqref{eqna9} into Eq.~\eqref{eqna5} gives
\begin{equation}
    \langle\{\hat{x},\hat{p}\}\rangle = 0.
    \label{eqna10}
\end{equation}

Since the modified Bessel function of the second kind admits the integral representation 
\begin{equation}
    K_{\nu}(x) = \int_0^\infty e^{-x\cosh t}\cosh(\nu t)\, dt.
    \label{eqna11}
\end{equation}
Differentiating Eq.~\eqref{eqna11} twice with respect to $\nu$ gives
\begin{equation}
    \frac{\partial^2 K_{\nu}(x)}{\partial \nu^2} = \int_0^\infty t^2\, e^{-x\cosh t}\, \cosh(\nu t)\, dt.
    \label{eqna12}
\end{equation}
For $\nu=0$ and $x=A$, the integral on the right coincides with that in Eq.~\eqref{eqna3}, and therefore
\begin{equation}
    \langle \hat{x}^2 \rangle = \frac{1}{\alpha^2K_0(A)} \left.\frac{\partial^2 K_{\nu}(A)}{\partial \nu^2}\right|_{\nu=0}.
    \label{eqna13}
\end{equation}
Equation~\eqref{eqna13} involves a derivative of $K_\nu$ with respect to its order rather than its argument; while closed-form expressions are not standard in the literature, this representation is well suited to stable numerical evaluation (or, alternatively, one may compute $\sigma_{11}$ via Gauss--Legendre quadrature).

\medskip
The use of differentiation under the integral sign is justified by Lebesgue's dominated convergence theorem~\cite{schilling2017measures}.\\

\bibliography{references_merged}% common bib file

\end{document}